\documentclass[multphys,vecphys]{svmult}
\usepackage{makeidx}     
\usepackage{graphicx}    
\usepackage{multicol}    

\makeindex             

\begin{document}

\title*{Tests for supernova explosion models: from light curves
to X-ray emission of supernova remnants}
\titlerunning{Tests for supernova explosion models}
\author{Elena Sorokina\inst{1}\and
        Sergey Blinnikov\inst{2}}
\institute{SAI, Universitetskij pr. 13, 119992 Moscow, Russia
\texttt{sorokina@sai.msu.su}
\and ITEP, B. Cheremushkinskaya 25, 117218 Moscow, Russia
\texttt{blinn@sai.msu.su}}
\maketitle

\section{Introduction}

So far, there exist many explosion models
proposed by theorists for dif\-fer\-ent types of supernovae,
but still there are no definite criteria to decide which
of the models are realized in nature.
Only a few parameters,
such as kinetic energy and total $^{56}$Ni
production, can be derived directly from the modelling of the explosion
and compared with the observational values.
The subsequent evolution of the exploded star gives us
much more possibilities to compare models and to decide which one fits
observations better.

The first possibility is modelling of $\gamma$-ray luminosity of supernovae.
The comparison with the observational values allows to define 
the total mass of radioactive isotopes, to judge on the composition 
of the outer layers where $\gamma$-rays are absorbed and thermalized, 
and also to check the approximations of $\gamma$-ray opacity.
During the first months after an explosion one can examine a theoretical model
by calculating bolometric and mono\-chro\-ma\-tic light curves and spectra,
and comparing them with observations.
Later on, gas in the ejecta cools down and becomes almost
unobservable. The next opportunity to analyze the ejecta is
on the stage of a young supernova remnant (SNR), when noticeable amount of
circumstellar gas is swept up. Then a reverse shock
forms, goes inwards the ejecta and illuminates them once again.

The successful theoretical supernova explosion models should be able
to explain any feature of the emission from supernovae 
at any evolutionary stage.

Different combinations of codes we have in our group allows us 
to compute the evolution of models, and therefore test them, 
on several stages.
For calculations of broad-band optical and bolometric light curves 
we use the multi-frequency radiation hydro code STELLA.
At the young remnant stage, while gas is transparent, we calculate 
the evolution and X-ray emission by combination of hydro part of STELLA 
with the non-equilibrium ionization code which is
based on the original algorithm by Peter Lundqvist.

\section{SN~Ia models}

We choose five models of Type Ia supernovae which are shown
in Table~\ref{sor_tab_models}.
W7~\cite{sor_w7} and DD4~\cite{sor_dd4} are more or less similar classical 1D 
Chandrasekhar mass models. 
LA4~\cite{sor_la4} and WD065~\cite{sor_wd065} are sub-Chandrasekhar mass ones.
(See original works for details.)
The last model was computed a couple of years ago in Garching, 
in the Max-Planck-Institute group~\cite{sor_mr}. It is originally 3D.
We average it over $4 \pi$ for our calculation, 
since we use only 1D codes at the moment.
The main feature of the model is that it is very well mixed, 
while energetics and the amount of Ni$^{56}$ are lower
than in other Chandrasekhar mass models.
In our calculations we compare two versions of the model with the same hydro 
part, but different element distributions.

\begin{table}
\centering
\caption{Parameters of SN Ia models}
\label{sor_tab_models}
\begin{tabular}{llllll}
\hline\noalign{\smallskip}
Model & DD4 & W7 & LA4 & WD065 & MR \\
\noalign{\smallskip}\hline\noalign{\smallskip}
$M_{\rm WD}{}^{\rm a}$      & 1.3861 & 1.3775 & 0.8678 & 0.6500  & 1.4 \\
$M_{{}^{56}{\rm Ni}}{}^{\rm a}$ & 0.63 & 0.60 & 0.47   & 0.05  & 0.42 \\ 
$E_{51}{}^{\rm b}$ & 1.23   & 1.20   & 1.15   & 0.56  & 0.46\\
\noalign{\smallskip}\hline
\multicolumn{5}{l}{${}^{\rm a}$in $M_\odot$} \\
\multicolumn{5}{l}{${}^{\rm b}$in $10^{51}$~ergs~s${}^{-1}$}
\end{tabular}
\end{table}

\section{Results}

\subsection{SNe~Ia}

Now there appears more and more possibilities to obtain $\gamma$--ray 
light curves using modern $\gamma$--ray space observatories, 
like COMPTEL~\cite{sor_diehl}. 
Here we just compare what one would 
observe after the explosion of the models we have discussed.

\begin{figure}
\centering
\includegraphics[height=7cm]{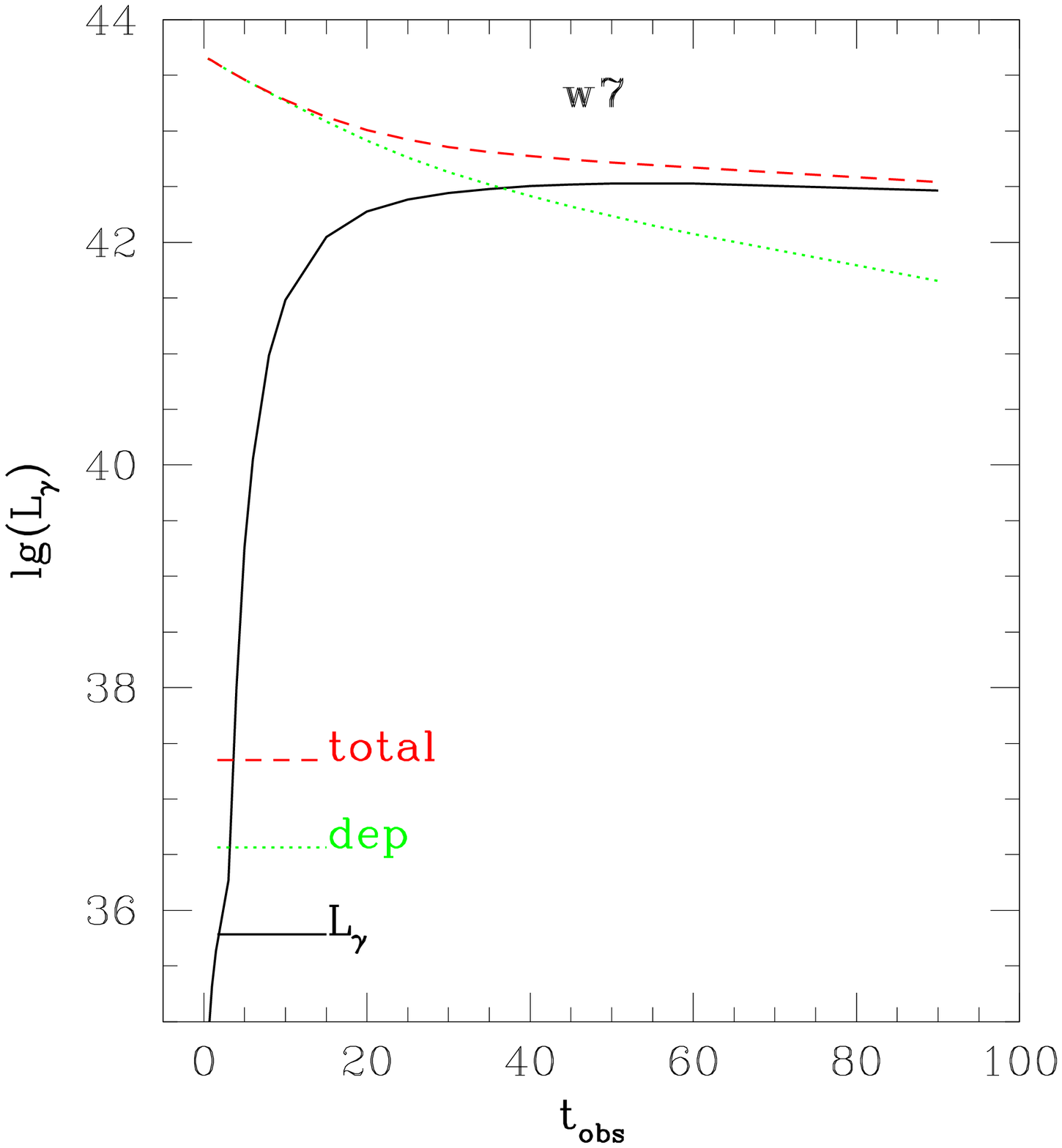}  \\
\includegraphics[height=7cm]{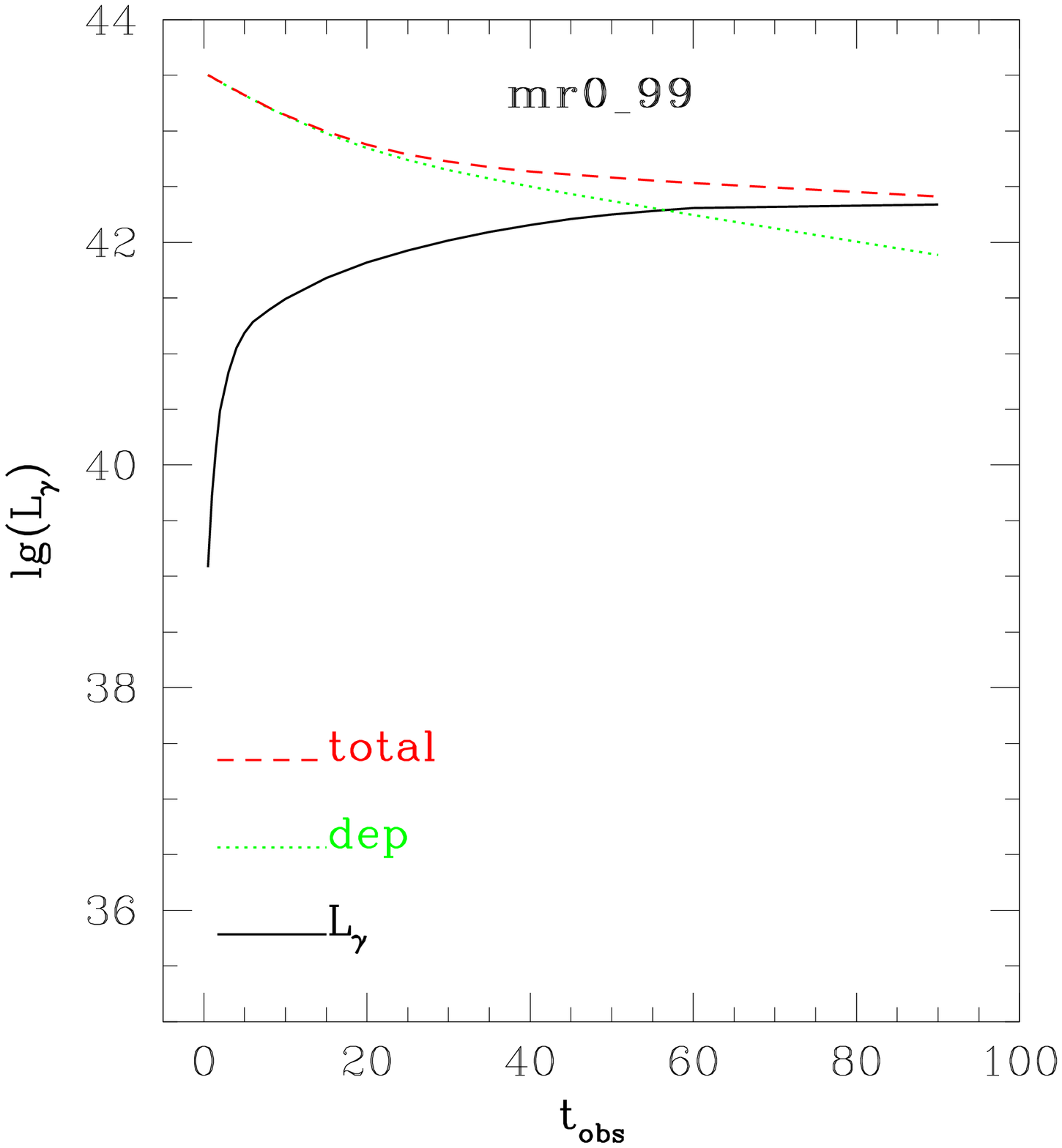}  
\includegraphics[height=7cm]{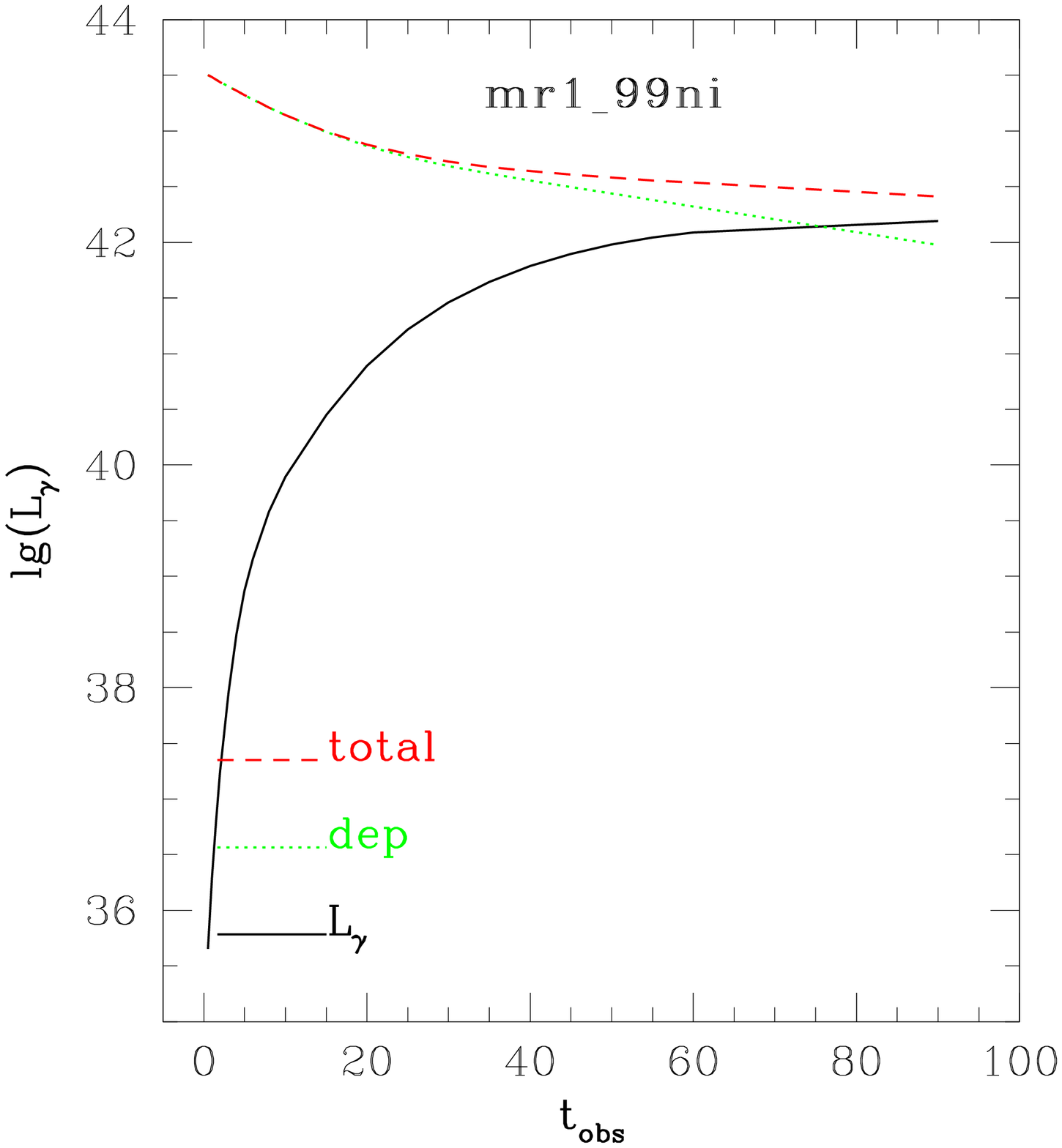}
\caption{$\gamma$--ray production by Ni and Co decays ({\it dashes}),
 $\gamma$--ray power deposited into heating ({\it dots}), and total
 $\gamma$--ray luminosity of SN ({\it solid}) versus time in days
 for different explosion models: W7 ({\it top panel}),
 MR model averaged over $4\pi$ ({\it bottom left}), and MR model averaged
 over an opening angle of $14^\circ$ with the same mass of ${}^{56}$Ni,
 which is situated mostly near the center ({\it bottom right}). }
\label{sor_fig_gam}
\end{figure}

In fig.~\ref{sor_fig_gam} we compare the light curves in $\gamma$--rays 
produced by different models. We have chosen W7 as a representative case of 
Chandrasekhar-mass models. The light curves in $\gamma$--rays are more or less
similar for all of them. The luminosity of Sub-Chandrasekhar-mass MR model is
lower due to smaller amount of ${}^{56}$Ni, and it rises slower since it is 
less energetic, so $\gamma$--photons are locked inside the ejecta 
for longer time. One can see even smaller number of $\gamma$--photons 
during the first weeks after the explosion in the model similar to MR 
but less mixed, with ${}^{56}$Ni residing mostly near the center. 
The maximum luminosity in $\gamma$ is still the same as in original MR, since 
we preserve the total amount of ${}^{56}$Ni.

Therefore, the light curve in $\gamma$--rays during the first months can be 
divided into two epochs, that represent different physical parameters 
of the explosion: the first 40--60 days and the following evolution.
From the observations during the first period one can judge on the combination
of explosion energy and the distribution of radioactive stuff over the ejecta,
while the second period tells us mostly about total production 
of radioactive elements at the explosion.
Since currently the observations can only provide us with fluxes 
integrated over several days (or even weeks), it seems expedient to make 
two different sets of observations: during the first two months 
after the explosion, and after that, in order to distinguish 
between different parameters of explosion, and to test explosion models.

We have described our modelling of bolometric and UBVRI light curves and 
calculations of X-ray emission of young SNRs in the previous work. 
For details one can see e.g. \cite{sor_porto,sor_uv,sor_tychocospar}. 
We just mention here that a bit less energetic and more mixed models, like MR, 
seem to us more preferable both for SN light curve and for SNR 
X-ray emission. At the latter point, our results seem to contradict
the conclusions by~\cite{sor_badenes}, that iron lines 
from well mixed models at the SNR stage are too strong, 
while less mixed models give a better agreement with observations of Tycho.
Still we trust in our results more, since we take into account 
the ionization energy in the equation of state, which is comparable
to the thermal energy. Electron 
thermal conductivity is also included into our calculations, and it smoothes
very much the temperature profile between forward and reverse shocks. 
Both these effects are neglected in the code by~\cite{sor_badenes}, though 
they are able to change the emitted spectrum appreciably.
The code by~\cite{sor_badenes} takes into account possible difference between 
electron and ion temperature using the standard Coulomb collision equation, 
while we treat this effect parametrically. This 
can also lead to differences in our results. The work on improving physics and
making our code more self-consistent is in progress.

\subsection{SN~1993J}

In order to pay attention to the SN entitling this conference, 
we have recalculated its light curve. We use the same model 13C 
from~\cite{sor_13C} which was investigated a few years ago 
in~\cite{sor_blineast93j}. It is one of the best models for SN~1993J, and we 
just wanted to check if the new version of our code {\sc STELLA}, 
with improved and renovated physics, still produces the light curve that fits 
observations well.

\begin{figure}
\centering
\includegraphics[height=7cm]{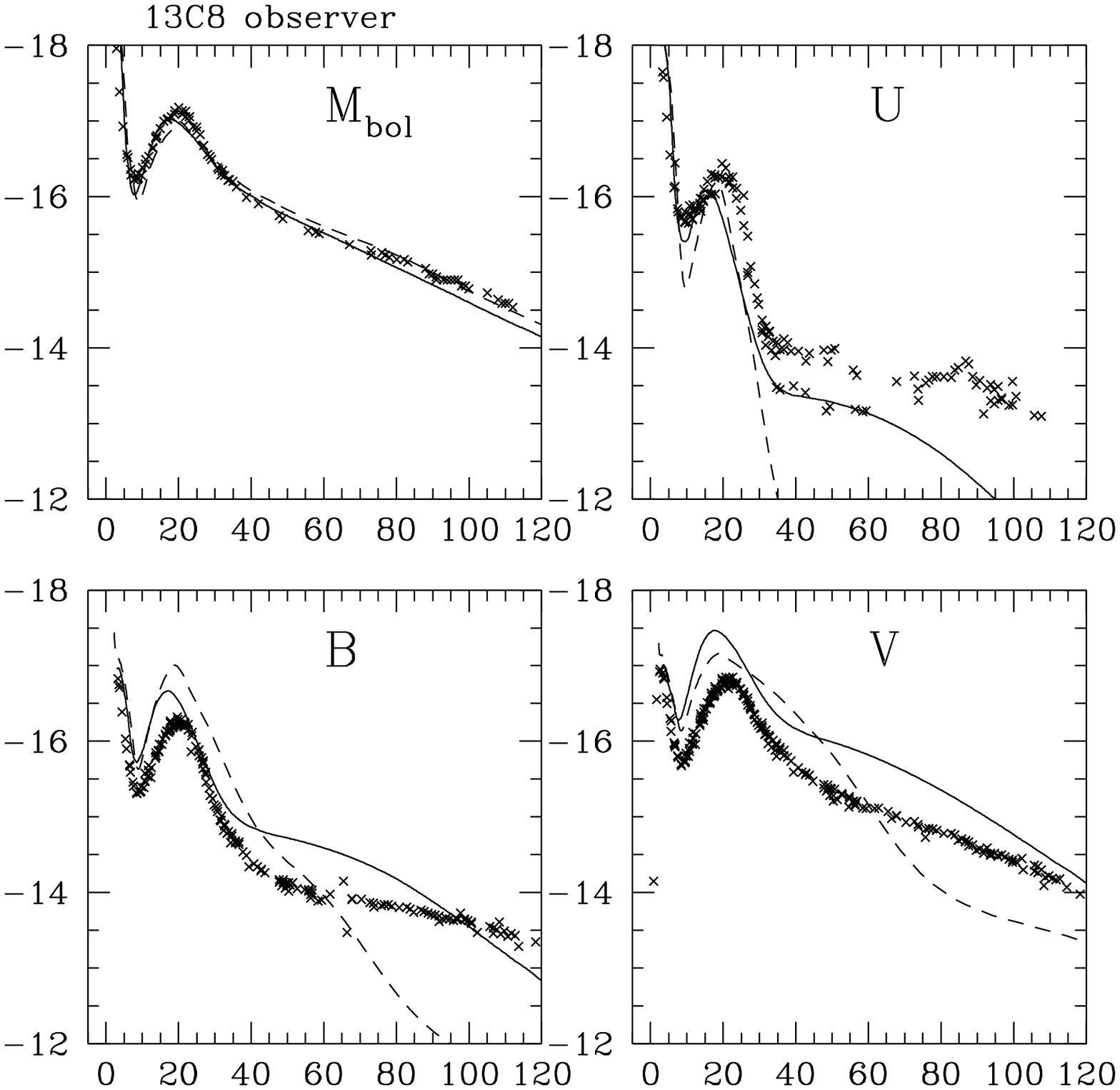}  
\includegraphics[height=7cm]{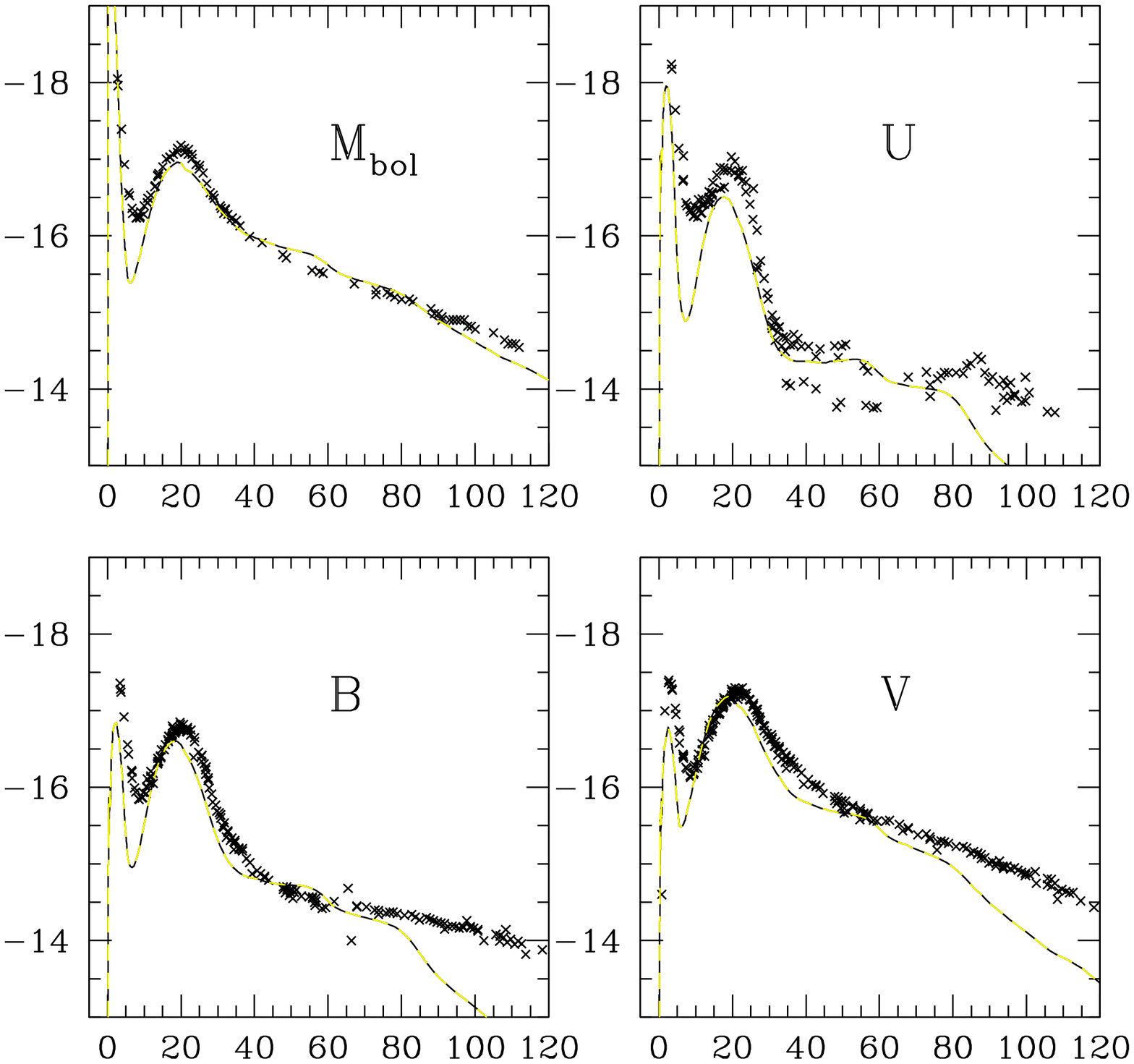}\\
\includegraphics[height=7cm]{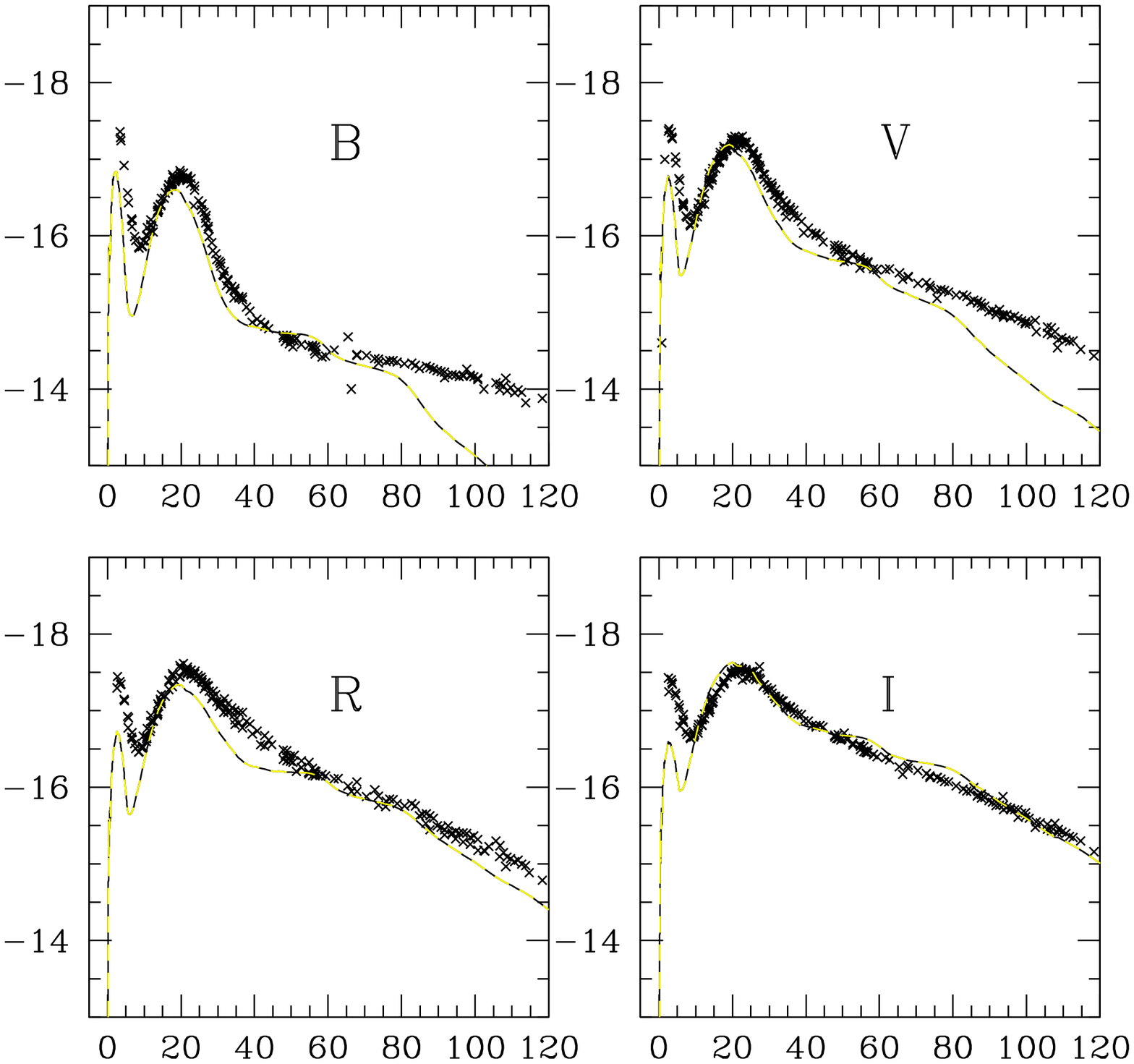}
\caption{Bolometric and UBVRI light curves for SN~1993J. {\it Top left panel} 
 shows the best model 13C8 from~\protect\cite{sor_blineast93j}
 ({\it solid line} -- calculations with {\sc STELLA}, {\it dashes} --
 with {\sc EDDINGTON}), {\it top right and bottom panels: } our recalculation
 of the same model with a new version of {\sc STELLA}. Observational data
 are taken from~\protect\cite{sor_richmond} and shown by crosses.
 The value of reddening is taken to be $E(B-V)=0.08$. }
\label{sor_fig_93j}
\end{figure}

The results of previous and current calculations are compared 
in fig.~\ref{sor_fig_93j}. The main improvement in the code is a new approach 
for expansion opacities~\cite{sor_expop}. They become much more complicated,
and, most probably, the bumps on the new light curves are the results of this 
improvement. But sometimes (for instance, in the U band) these bumps seem to 
fit the observations even better than it was in the old version. R and I bands 
are calculated with {\sc STELLA} for the first time, and they look perfect. 

There is a discussion in the literature on the estimates of the reddening 
to SN~1993J (\cite{sor_richmond} and references therein). In the plots 
in fig.~\ref{sor_fig_93j} we assume $E(B-V)=0.08$. With larger value, 
the observational curves become higher than the modelled ones. 
So the new version of {\sc STELLA} confirms that the model 13C corresponds 
to the explosion of SN~1993J very well, but in the case of low reddening.

{\small
{\it Acknowledgements.} ES is grateful to the organizers of the meeting 
for their warm hospitality and support. The work is supported by RFBR through 
grants 02--02--16500, 03--02--06770, and 03--02--26598.
}


\printindex

\begin{thebibliography}{99.}

\bibitem{sor_w7} K.~Nomoto, F.--K.~Thielemann, K.~Yokoi: ApJ, {\bf 286}, 644 
   (1984)

\bibitem{sor_dd4} S.E.~Woosley, T.A.~Weaver: Massive Stars, Supernovae, 
   and Nucleosynthesis. In: \textit{Supernovae}, ed by R.~Bludman et al 
   (ELSEVIER Sci. Pub., Amsterdam 1994) p 63--154

\bibitem{sor_la4} E.~Livne, D.~Arnett: ApJ \textbf{452}, 62 (1995)

\bibitem{sor_wd065} P.~Ruiz-Lapuente et al.: Nature \textbf{365}, 728 (1993)

\bibitem{sor_mr} M.~Reinecke, W.~Hillebrandt,  J.~C.~Niemeyer:  A\&A 
   \textbf{386}, 936 (2002)

\bibitem{sor_diehl} R.~Georgii, S.~Pl\"uschke, R.~Diehl et al: A\&A
   \textbf{394},517 (2002)

\bibitem{sor_porto} S.~Blinnikov, E.~Sorokina: Type Ia Supernova models: 
   latest developments. In: \textit{Hunting the cosmological parameters},
   ed by D.~Barbossa et al (KLUWER 2003) (astro--ph/0212530)

\bibitem{sor_uv} S.~Blinnikov, E.~Sorokina: A\&A \textbf{356}, L30 (2000)

\bibitem{sor_tychocospar} D.I.~Kosenko, E.I.Sorokina, S.I.~Blinnikov,
   P.~Lundqvist: X-ray emission of young SN~Ia remnants as a probe 
   for an explosion model. In: \textit{Proc. of 34th Scientific Assembly
    of COSPAR}, (ASR 2003) (astro-ph/0212188)

\bibitem{sor_badenes} C.~Badenes, E.~Bravo, K.J.~Borkowski,
   I.~Dom$\acute\imath$nguez: ApJ \textbf{593} 358 (2003)

\bibitem{sor_13C} S.E.~Woosley, R.G.~Eastman, T.A.~Weaver, P.A.~Pinto: ApJ 
   \textbf{429}, 300 (1994)

\bibitem{sor_blineast93j} S.I.~Blinnikov, R.~Eastman, O.S.~Bartunov et al: ApJ
   \textbf{496}, 454 (1998)

\bibitem{sor_expop} E.I.~Sorokina, S.I.~Blinnikov: Energy exchange
   inside SN ejecta and light curves of SNe~Ia.
   In: \textit{Nuclear Astrophysics}, MPA/P13, ed by W.~Hillebrandt,
   E.~M\"uller (Garching 2002) pp 57--62

\bibitem{sor_richmond} M.W.~Richmond, R.R.~Treffers, A.V.~Filippenko et al:
   Astron. J. \textbf{107}, 1022 (1994)





\end{thebibliography}
\end{document}